\documentclass[USenglish,twocolumn]{article}

\usepackage[utf8]{inputenc}
\usepackage[big]{dgruyter}
\usepackage{microtype}
\usepackage{color}
\usepackage{algorithm}

\usepackage{wrapfig}

\usepackage{mathtools}
\usepackage{subfigure}

\newtheorem{theorem}{\bf Theorem} \newtheorem{definition}{\bf Definition}

\newtheorem{Algorithm}{\bf Algorithm}

\usepackage{setspace}
\begin{document}
  \articletype{Methods}

  \author*[1]{Julian Berberich}
  \author[2]{Johannes K\"ohler}
  \author[3]{Matthias A. M\"uller}
  \author[2]{Frank Allg\"ower} 
  \runningauthor{J. Berberich, J. K\"ohler, M. A. M\"uller, F. Allg\"ower}
  \affil[1]{Universit\"at Stuttgart, Institut f\"ur Systemtheorie und Regelungstechnik, 70550 Stuttgart}
  \affil[2]{Universit\"at Stuttgart, Institut f\"ur Systemtheorie und Regelungstechnik, 70550 Stuttgart}
  \affil[3]{Leibniz Universit\"at Hannover, Institut f\"ur Regelungstechnik, 30167 Hannover\newline
  \copyright 2021 Berberich et al., published by De Gruyter. This work is licensed under the Creative Commons Attribution 4.0 International License (CC-BY).}
  \title{Data-driven model predictive control: closed-loop guarantees and experimental results}
  \runningtitle{Data-driven model predictive control}
  \subtitle{Datenbasierte pr\"adiktive Regelung: theoretische Garantien und experimentelle Ergebnisse}
  \abstract{
  We provide a comprehensive review and practical implementation of a recently developed model predictive control (MPC) framework for controlling unknown systems using only measured data and no explicit model knowledge.
  Our approach relies on an implicit system parametrization from behavioral systems theory based on one measured input-output trajectory.
  The presented MPC schemes guarantee closed-loop stability for unknown linear time-invariant (LTI) systems, even if the data are affected by noise.
  Further, we extend this MPC framework to control unknown nonlinear systems by continuously updating the data-driven system representation using new measurements.
  The simple and intuitive applicability of our approach is demonstrated with a nonlinear four-tank system in simulation and in an experiment.\\
  \vskip2pt  
  \textbf{Zusammenfassung:}
  Dieser Artikel beinhaltet einen umfassenden Überblick sowie eine praktische Implementierung von kürzlich entwickelten Entwurfsverfahren zur modellprädiktiven Regelung (MPC), welche unbekannte Systeme nur mit Hilfe von gemessenen Daten und ohne explizites Modellwissen regeln.
  Unser Ansatz bedient sich einer impliziten Systemparametrisierung aus der \emph{behavioral} Systemtheorie basierend auf einer Eingangs-Ausgangs-Trajektorie.
  Die präsentierten MPC-Algorithmen garantieren Stabilität für unbekannte lineare, zeitinvariante Systeme, selbst im Fall von verrauschten Messungen.
  Zusätzlich stellen wir eine Erweiterung vor, um unbekannte nichtlineare Systeme zu regeln durch stetige Aktualisierung der datenbasierten Systemparametrisierung.
  Die einfache und intuitive Anwendbarkeit wird an einem nichtlinearen Vier-Tank System in der Simulation und in einem Experiment demonstriert.}
  \received{January 30, 2021}
  \accepted{May 17, 2021}
  \journalname{at-Automatisierungstechnik}
  \journalyear{2021}
  \journalvolume{69}
  \journalissue{7}
  \startpage{608}
  \DOI{10.1515/auto-2021-0024}

\maketitle

\section{Introduction}
Model predictive control (MPC) is a successful modern control technique which relies on the repeated solution of an open-loop optimal control problem~\cite{rawlings2017model}.
Essential advantages of MPC are its applicability to general system classes and the possibility to enforce constraint satisfaction.
In order to implement an MPC controller, typically an accurate model of the plant is required.
Since modeling is often the most time-consuming step in controller design and due to the increasing availability of data, control approaches using only data and inaccurate or no model knowledge have recently gained increasing attention~\cite{hou2013model}.
Examples for such approaches are recent works on adaptive~\cite{adetola2011robust,aswani2013provably} or learning-based~\cite{hewing2020learning} MPC.

Another promising approach for designing MPC schemes using only measured data stems from a result from behavioral systems theory:
In~\cite{willems2005note}, it is shown that one input-output trajectory of an unknown linear time-invariant (LTI) system can be used to parametrize all trajectories, assuming that the corresponding input is persistently exciting.
By replacing the standard state-space model with this data-dependent parametrization, it is simple to design MPC schemes which use input-output data instead of prior model knowledge~\cite{yang2015data,coulson2019deepc,berberich2021guarantees}.
Such MPC schemes have successfully been applied to challenging real-world examples, compare~\cite{elokda2019quadcopters}, and open-loop robustness properties have been established~\cite{coulson2020distributionally}.
However, for a reliable application to complex or safety-critical systems, guarantees for the closed-loop behavior are crucial, which are, however, challenging to obtain, in particular in case of noisy data.

In this paper, we provide an overview of recent advances in data-driven MPC based on~\cite{willems2005note}.
We focus on MPC schemes with guaranteed closed-loop stability and robustness properties in case of LTI systems~\cite{berberich2021guarantees,berberich2020tracking,berberich2020constraints,berberich2021on,bongard2021robust}.
Additionally, we demonstrate how such MPC schemes can be modified to control unknown nonlinear systems using only measured data.
We perform an extensive validation of this approach in simulation and in an experiment involving the classical nonlinear four-tank system from~\cite{raff2006nonlinear}.

The remainder of the paper is structured as follows.
After providing some preliminaries in Section~\ref{sec:prelim}, we present MPC schemes to control LTI systems using noise-free data, LTI systems using noisy data, and nonlinear systems, respectively, in Section~\ref{sec:mpc}.
We then validate the presented MPC framework with a nonlinear four-tank system in simulation (Section~\ref{sec:simulation}) and in an experiment (Section~\ref{sec:experiment}).
Finally, we conclude the paper in Section~\ref{sec:conclusion}.

\section{Preliminaries}\label{sec:prelim}
We write $\mathbb{I}_{[a,b]}$ for the set of all integers in the interval $[a,b]$, $\mathbb{I}_{\geq0}$ for the set of nonnegative integers, and $\mathbb{R}_{\geq0}$ for the set of nonnegative real numbers.
For a vector $x$, we denote by $\lVert x\rVert_p$ its $p$-norm.
We denote an identity matrix of appropriate dimension by $I$, we write $P=P^\top\succ0$ if a matrix $P$ is positive definite, and we define $\lVert x\rVert_P^2\coloneqq x^\top Px$.
The interior of a set $X$ is denoted by $\mathrm{int}(X)$.
We define $\mathcal{K}$ as the class of functions $\alpha:\mathbb{R}_{\geq0}\to\mathbb{R}_{\geq0}$ which are continuous, strictly increasing, and satisfy $\alpha(0)=0$.
For a sequence $\{u_k\}_{k=0}^{N-1}$, we define the Hankel matrix
\begin{align*}
H_L(u)\coloneqq\begin{bmatrix}u_0&u_1&\dots&u_{N-L}\\
u_1&u_2&\dots&u_{N-L+1}\\
\vdots&\vdots&\ddots&\vdots\\
u_{L-1}&u_L&\dots&u_{N-1}\end{bmatrix}
\end{align*}
and we write $u_{[a,b]}\coloneqq\begin{bmatrix}u_a^\top&\dots&u_b^\top\end{bmatrix}^\top$, $u\coloneqq u_{[0,N-1]}$.
For our theoretical results, we consider an LTI system
\begin{align}\label{eq:sys}
x_{k+1}&=Ax_k+Bu_k,\>\>
y_k=Cx_k+Du_k
\end{align}
with state $x_k\in\mathbb{R}^n$, input $u_k\in\mathbb{R}^m$, and output $y_k\in\mathbb{R}^p$.
Throughout this paper, we make the standing assumption that $(A,B)$ is controllable, $(A,C)$ is observable, and an upper bound on the system order $n$ is known. 
Beyond that, no knowledge on System~\eqref{eq:sys} is available and, in particular, the matrices $A$, $B$, $C$, $D$ are unknown.
A measured input-output trajectory $\{u_k^d,y_k^d\}_{k=0}^{N-1}$ is assumed to be available, where the input $u^d$ is persistently exciting.
\begin{definition}\label{def:pe}
We say that a sequence $\{u_k\}_{k=0}^{N-1}$ with $u_k\in\mathbb{R}^m$ is persistently exciting of order $L$ if $\mathrm{rank}(H_L(u))=mL$.
\end{definition}
Note that persistence of excitation of order $L$ imposes a lower bound on the required data length $N$, i.e., $N\geq(m+1)L-1$.
The following result provides a purely data-driven parametrization of all trajectories of~\eqref{eq:sys}.
While the result is originally formulated and proven in the behavioral framework in~\cite{willems2005note}, we state a reformulation in the state-space framework from~\cite{berberich2020trajectory}.
\begin{theorem}\label{thm:hankel}
(\cite[Theorem 3]{berberich2020trajectory})
Suppose $\{u_k^d,y_k^d\}_{k=0}^{N-1}$ is a trajectory of~\eqref{eq:sys}, where $u^d$ is persistently exciting of order $L+n$.
Then, $\{\bar{u}_k,\bar{y}_k\}_{k=0}^{L-1}$ is a trajectory of~\eqref{eq:sys} if and only if there exists $\alpha\in\mathbb{R}^{N-L+1}$ such that
\begin{align}\label{eq:thm_hankel}
\begin{bmatrix}H_L(u^d)\\H_L(y^d)\end{bmatrix}\alpha=\begin{bmatrix}\bar{u}\\\bar{y}\end{bmatrix}.
\end{align}
\end{theorem}
Theorem~\ref{thm:hankel} shows that Hankel matrices containing one persistently exciting input-output trajectory span the space of all system trajectories.
This allows us to parametrize any trajectory of an unknown system, using only measured data and no explicit model knowledge.
While verifying the condition on $u^d$ in Theorem~\ref{thm:hankel} requires knowledge of the system order $n$, the result (and all further results in this paper relying on Theorem~\ref{thm:hankel}) remains true if $n$ is replaced by a (potentially rough) upper bound. 

\section{Data-driven model predictive control}\label{sec:mpc}
In this section, we review data-driven MPC schemes based on Theorem~\ref{thm:hankel} with a special focus on the closed-loop guarantees that can be given for such schemes if applied to LTI systems.
We address the cases of noise-free data (Section~\ref{subsec:mpc_nominal}) and noisy data (Section~\ref{subsec:mpc_robust}) both for LTI systems.
Furthermore, we present a data-driven MPC scheme to control nonlinear systems in Section~\ref{subsec:mpc_nl}.

\subsection{Nominal data-driven MPC for LTI systems}\label{subsec:mpc_nominal}
Our goal is to track a given input-output setpoint $(u^s,y^s)\in\mathbb{U}\times\mathbb{Y}$ which corresponds to an equilibrium of the system~\eqref{eq:sys}, i.e., $\{u_k,y_k\}_{k=0}^{n}$ with $(u_k,y_k)=(u^s,y^s)$, $k\in\mathbb{I}_{[0,n]}$ is a valid trajectory of~\eqref{eq:sys} (compare~\cite[Definition 3]{berberich2021guarantees}).
At the same time, we want to satisfy pointwise-in-time constraints $u_t\in\mathbb{U}$, $y_t\in\mathbb{Y}$ for given constraint sets $\mathbb{U}\subseteq\mathbb{R}^m$, $\mathbb{Y}\subseteq\mathbb{R}^p$.
MPC is a well-established method which can be used to achieve this task.
It relies on the repeated solution of an open-loop optimal control problem, optimizing over all possible future system trajectories at each time step and always applying the first input component~\cite{rawlings2017model}.
Standard MPC approaches exploit model knowledge, i.e., knowledge of the matrices $A$, $B$, $C$, $D$ in~\eqref{eq:sys}, in order to solve this optimization problem.
In contrast, the MPC scheme we consider relies on Theorem~\ref{thm:hankel} which parametrizes all possible system trajectories, using only one input-output trajectory $\{u_k^d,y_k^d\}_{k=0}^{N-1}$.

\begin{subequations}\label{eq:DD_MPC}
Future trajectories can only be uniquely predicted if an additional initial condition is imposed, compare~\cite{markovsky2008data}.
Therefore, since we assume that only input-output data of~\eqref{eq:sys} and no state measurements are available, we use the last $n$ input-output measurements $\{u_k,y_k\}_{k=t-n}^{t-1}$ to implicitly specify initial conditions at time $t$ and thus, to fix a unique system trajectory.
Based on these ingredients, we define the following optimal control problem:
\begin{align}
\underset{\substack{\alpha(t),\bar{u}(t),\bar{y}(t)}}{\min}\>\>&\sum_{k=0}^{L-1}
\lVert\bar{u}_k(t)-u^s\rVert_R^2+\lVert\bar{y}_k(t)-y^s\rVert_Q^2\\\label{eq:DD_MPC_model}
\text{s.t.}&\begin{bmatrix}\bar{u}_{[-n,L-1]}(t)\\\bar{y}_{[-n,L-1]}(t)\end{bmatrix}=\begin{bmatrix}H_{L+n}(u^d)\\H_{L+n}(y^d)\end{bmatrix}\alpha(t),\\
\label{eq:DD_MPC_init}
&\begin{bmatrix}\bar{u}_{[-n,-1]}(t)\\\bar{y}_{[-n,-1]}(t)\end{bmatrix}=\begin{bmatrix}u_{[t-n,t-1]}\\y_{[t-n,t-1]}\end{bmatrix},\\\label{eq:DD_MPC_tec}
&\begin{bmatrix}\bar{u}_{[L-n,L-1]}(t)\\\bar{y}_{[L-n,L-1]}(t)\end{bmatrix}=\begin{bmatrix}u^s_n\\y^s_n\end{bmatrix},\\\label{eq:DD_MPC_constraints}
&\bar{u}_k(t)\in\mathbb{U},\>\>\bar{y}_k(t)\in\mathbb{Y},\>\>k\in\mathbb{I}_{[0,L-1]}.
\end{align}
\end{subequations}
Problem~\eqref{eq:DD_MPC} takes a common MPC form, minimizing the difference of the predicted input-output variables $\bar{u}(t)$, $\bar{y}(t)$ w.r.t. the setpoint $(u^s,y^s)$ while satisfying the constraints in~\eqref{eq:DD_MPC_constraints}.
The matrices $Q,R\succ0$ are weights for tuning which can be specified by the user.
The key difference to standard model-based MPC is that the ``prediction model'' is formed based on Theorem~\ref{thm:hankel}, i.e., by using Hankel matrices in~\eqref{eq:DD_MPC_model}.
Moreover,~\eqref{eq:DD_MPC_init} initializes the predictions using the last $n$ input-output measurements, which implies that the internal states of the predictions and of the system at time $t$ coincide.
Due to these initial conditions, the predictions have an overall length of $L+n$.

Further, the constraint~\eqref{eq:DD_MPC_tec} is a \emph{terminal equality constraint} on the last $n$ input-output predictions, similar to model-based MPC~\cite{rawlings2017model}, where such conditions can be imposed on the state to ensure closed-loop stability.
In Equation~\eqref{eq:DD_MPC_tec}, we write $u^s_n$, $y^s_n$ for column vectors containing $n$ times $u^s$ and $y^s$, respectively.
The constraint~\eqref{eq:DD_MPC_tec} is the main difference of Problem~\eqref{eq:DD_MPC} to other works on data-driven MPC, e.g., in~\cite{coulson2019deepc,yang2015data}, and it can be used to prove closed-loop stability for the presented MPC scheme.
Note that Problem~\eqref{eq:DD_MPC} does not require offline or online state measurements and hence, the considered MPC approach is inherently an output-feedback MPC.

For polytopic constraints, Problem~\eqref{eq:DD_MPC} is a convex quadratic program (QP) which can be solved efficiently, similar to model-based MPC.
Throughout this section, we write $u_t$, $x_t$, $y_t$ for closed-loop variables at time $t\in\mathbb{I}_{\geq0}$, and $\{\bar{u}_k^*(t),\bar{y}_k^*(t)\}_{k=-n}^{L-1}$ for the optimal solution predicted at time $t$.
Problem~\eqref{eq:DD_MPC} is applied in a standard receding horizon fashion which is summarized in Algorithm~\ref{alg:MPC_nom}.
\begin{algorithm}
\begin{Algorithm}\label{alg:MPC_nom}
\normalfont{\textbf{Nominal Data-Driven MPC}}\\
\normalfont{
\textbf{Offline:} Choose upper bound on system order $n$, prediction horizon $L$, cost matrices $Q,R\succ0$, constraint sets $\mathbb{U},\mathbb{Y}$, setpoint $(u^s,y^s)$, and generate data $\{u_k^d,y_k^d\}_{k=0}^{N-1}$.}\\
\normalfont{\textbf{Online:}}
\begin{enumerate}
\item At time $t$, take the past $n$ measurements $\{u_k,y_k\}_{k=t-n}^{t-1}$ and solve~\eqref{eq:DD_MPC}.
\item Apply the input $u_t=\bar{u}_0^*(t)$.
\item Set $t=t+1$ and go back to 1).
\end{enumerate}
\end{Algorithm}
\end{algorithm}

The following result summarizes the closed-loop properties of Algorithm~\ref{alg:MPC_nom} when applied to~\eqref{eq:sys}.

\begin{theorem}\label{thm:DD_MPC_nom}
(\cite[Theorem 2]{berberich2021guarantees})
Suppose $L\geq n$, $u^d$ is persistently exciting of order $L+2n$, and the optimal cost of~\eqref{eq:DD_MPC} is upper bounded by\footnote{We define $x^s$ as the steady-state corresponding to $(u^s,y^s)$.} $c_u\lVert x_t-x^s\rVert_2^2$ for some $c_u>0$~\cite[Assumption 1]{berberich2021guarantees}.
If Problem~\eqref{eq:DD_MPC} is feasible at $t=0$, then
\begin{itemize}
\item it is feasible at any $t\in\mathbb{I}_{\geq0}$,
\item the closed loop satisfies the constraints, i.e., $u_t\in\mathbb{U}$ and $y_t\in\mathbb{Y}$ for all $t\in\mathbb{I}_{\geq0}$,
\item the steady-state $x^s$ is exponentially stable for the resulting closed loop.
\end{itemize}
\end{theorem} 

Theorem~\ref{thm:DD_MPC_nom} shows that the simple MPC scheme based on repeatedly solving~\eqref{eq:DD_MPC} stabilizes the unknown LTI system~\eqref{eq:sys}, using only one a priori collected input-output trajectory.
The proof is similar to stability arguments in model-based MPC~\cite{rawlings2017model} with the additional difficulty that the cost of~\eqref{eq:DD_MPC} depends on the output and is thus only positive \emph{semi}-definite in the internal state.
The assumption that the cost of~\eqref{eq:DD_MPC} is quadratically upper bounded is not restrictive and it holds, e.g., for compact constraints if $(u^s,y^s)\in\mathrm{int}(\mathbb{U}\times\mathbb{Y})$ (see~\cite{berberich2021guarantees,berberich2021on} for details).

Since Theorem~\ref{thm:hankel} provides an equivalent parametrization of system trajectories, its applicability is not limited to MPC schemes with terminal equality constraints as above.
In particular, it can be used to design more sophisticated MPC schemes with general terminal ingredients, i.e., a terminal cost and a terminal region constraint, see~\cite{berberich2021on} for details.
Similar to terminal ingredients in model-based MPC, this has the advantage of increasing the region of attraction and improving robustness in closed loop.
Alternatively, Theorem~\ref{thm:hankel} is used to design a data-driven tracking MPC scheme in~\cite{berberich2020tracking}, where the setpoint $(u^s,y^s)$ for which the terminal equality constraint~\eqref{eq:DD_MPC_tec} is imposed is optimized online, analogously to model-based tracking MPC~\cite{limon2008mpc}.
In the data-driven problem setting considered in this paper, such a tracking formulation has the advantage that the given input-output setpoint need not be an equilibrium of the \emph{unknown} system~\eqref{eq:sys}, which is a property that may be difficult to verify in practice.
Finally,~\cite{bongard2021robust} provides closed-loop stability and robustness guarantees for a data-driven MPC scheme \emph{without any terminal ingredients} for both noise-free and noisy input-output data.

\subsection{Robust data-driven MPC for LTI systems}\label{subsec:mpc_robust}
Theorem~\ref{thm:DD_MPC_nom} only applies if the measured data are noise-free, which is rarely the case in a practical application.
In this section, we consider the more challenging case of noisy data.
In particular, we assume that both the data used for prediction as well as the initial conditions are affected by bounded output measurement noise, i.e., we have access to $\{u_k^d,\tilde{y}_k^d\}_{k=0}^{N-1}$ and $\{u_k,\tilde{y}_k\}_{k=t-n}^{t-1}$, where $\tilde{y}_k^d=y_k^d+\varepsilon_k^d$ and $\tilde{y}_k=y_k+\varepsilon_k$ with the noise satisfying the bound $\lVert\varepsilon_k^d\rVert_{\infty}\leq\bar{\varepsilon}$, $\lVert\varepsilon_k\rVert_{\infty}\leq\bar{\varepsilon}$ for $k\in\mathbb{I}_{\geq0}$ for some $\bar{\varepsilon}>0$.
In order to retain desirable closed-loop properties despite noisy measurements, we consider the following modified data-driven MPC scheme:
\begin{subequations}\label{eq:DD_MPC_robust}
\begin{align}\label{eq:DD_MPC_robust_cost}
\underset{\substack{\alpha(t),\sigma(t)\\\bar{u}(t),\bar{y}(t)}}{\min}&\sum_{k=0}^{L-1}
\lVert\bar{u}_k(t)-u^s\rVert_R^2+\lVert\bar{y}_k(t)-y^s\rVert_Q^2\\\nonumber
\qquad&+\lambda_\alpha\bar{\varepsilon}\lVert\alpha(t)\rVert_2^2+\frac{\lambda_\sigma}{\bar{\varepsilon}}\lVert\sigma(t)\rVert_2^2\\
\label{eq:DD_MPC_robust_model} \text{s.t.}\>\> &\>\begin{bmatrix}
\bar{u}(t)\\\bar{y}(t)+\sigma(t)\end{bmatrix}=\begin{bmatrix}H_{L+n}\left(u^d\right)\\H_{L+n}\left(\tilde{y}^d\right)\end{bmatrix}\alpha(t),\\\label{eq:DD_MPC_robust_init}
&\>\begin{bmatrix}\bar{u}_{[-n,-1]}(t)\\\bar{y}_{[-n,-1]}(t)\end{bmatrix}=\begin{bmatrix}u_{[t-n,t-1]}\\\tilde{y}_{[t-n,t-1]}\end{bmatrix},\\\label{eq:DD_MPC_robust_tec}
&\>\begin{bmatrix}\bar{u}_{[L-n,L-1]}(t)\\\bar{y}_{[L-n,L-1]}(t)\end{bmatrix}=\begin{bmatrix}u^s_n\\y^s_n\end{bmatrix},\>\bar{u}_k(t)\in\mathbb{U}.
\end{align}
\end{subequations}
In order to account for the noise affecting the available data in~\eqref{eq:DD_MPC_robust_model}, Problem~\eqref{eq:DD_MPC_robust} contains an additional slack variable $\sigma(t)$.
Both the slack variable and the vector $\alpha(t)$ are regularized in the cost, where the regularization depends on parameters $\lambda_{\alpha},\lambda_{\sigma}>0$ as well as on the noise level $\bar{\varepsilon}$.
The regularization of $\alpha(t)$ is needed since there exist infinitely many $\alpha$ satisfying~\eqref{eq:thm_hankel} for a given input-output trajectory.
The noise in the data $\tilde{y}^d$ acts as a multiplicative uncertainty w.r.t. $\alpha(t)$ in~\eqref{eq:DD_MPC_robust_model} and thus regularizing the norm of $\alpha(t)$ reduces the influence of the noise on the prediction accuracy.
On the other hand, the regularization of $\sigma(t)$ prevents large values of $\sigma(t)$ which may also deteriorate the prediction accuracy.
Note that Problem~\eqref{eq:DD_MPC_robust} recovers the nominal MPC scheme in Problem~\eqref{eq:DD_MPC} for $\bar{\varepsilon}\to0$.
In~\cite{berberich2021guarantees}, an additional (non-convex) constraint on $\sigma(t)$ was required, but it was recently shown in~\cite{bongard2021robust} that this constraint can be dropped if the regularization of $\sigma(t)$ depends reciprocally on $\bar{\varepsilon}$, cf.~\eqref{eq:DD_MPC_robust_cost}.
Hence, if $\mathbb{U}$ is a convex polytope, Problem~\eqref{eq:DD_MPC_robust} is a strictly convex QP.

For simplicity, we do not consider output constraints in~\eqref{eq:DD_MPC_robust}, i.e., $\mathbb{Y}=\mathbb{R}^p$.
It is possible to extend the presented results by including a constraint tightening which guarantees robust output constraint satisfaction despite output measurement noise, see~\cite{berberich2020constraints}.
Finally, we note that MPC schemes similar to Problem~\eqref{eq:DD_MPC_robust} have been proposed in~\cite{coulson2019deepc,coulson2020distributionally}, but only open-loop robustness properties have been proven.
In the following, we state closed-loop properties resulting from the application of Problem~\eqref{eq:DD_MPC_robust} in a multi-step fashion, see Algorithm~\ref{alg:MPC_rob}.
%
\begin{algorithm}
\begin{Algorithm}\label{alg:MPC_rob}
\normalfont{\textbf{Robust Data-Driven MPC}}\\
\normalfont{
\textbf{Offline:} Choose upper bound on system order $n$, prediction horizon $L$, cost matrices $Q,R\succ0$, regularization parameters $\lambda_{\alpha},\lambda_{\sigma}>0$, constraint set $\mathbb{U}$, noise bound $\bar{\varepsilon}>0$, setpoint $(u^s,y^s)$, and generate data $\{u_k^d,\tilde{y}_k^d\}_{k=0}^{N-1}$.}\\
\normalfont{\textbf{Online:}}
\begin{enumerate}
\item At time $t$, take the past $n$ measurements $\{u_k,\tilde{y}_k\}_{k=t-n}^{t-1}$ and solve~\eqref{eq:DD_MPC_robust}.
\item Over the next $n$ time steps, apply the input $u_{[t,t+n-1]}=\bar{u}_{[0,n-1]}^*(t)$.
\item Set $t=t+n$ and go back to 1).
\end{enumerate}
\end{Algorithm}
\end{algorithm}
We consider a multi-step MPC scheme due to the joint occurrence of model mismatch, i.e., output measurement noise in the Hankel matrix in~\eqref{eq:DD_MPC_robust_model}, and terminal equality constraints.
Due to this combination and the controllability argument used to prove stability in~\cite{berberich2021guarantees}, the theoretical guarantees are only valid locally for a one-step MPC scheme~\cite[Remark 4]{berberich2021guarantees}.
When removing the terminal equality constraints~\eqref{eq:DD_MPC_robust_tec} as in~\cite{bongard2021robust} or replacing them by general terminal ingredients~\cite{berberich2021on}, then comparable closed-loop guarantees can also be given for a one-step scheme.
\begin{theorem}\label{thm:DD_MPC_rob}
(\cite[Theorem 3]{berberich2021guarantees})
Suppose $L\geq2n$, $u^s=0\in\mathrm{int}(\mathbb{U})$, and $u^d$ is persistently exciting of order $L+2n$.
Then, there exist a set $\mathbb{X}_V\subseteq\mathbb{R}^n$, parameters $\lambda_{\alpha},\lambda_{\sigma}>0$, a sufficiently small noise bound $\bar{\varepsilon}>0$, and a function $\beta\in\mathcal{K}$ such that $\mathbb{X}_V$ is positively invariant and $x_t$ converges exponentially to $\{x\in\mathbb{R}^n\mid\lVert x\rVert_2\leq\beta(\bar{\varepsilon})\}$ in closed loop.
\end{theorem}

Theorem~\ref{thm:DD_MPC_rob} should be interpreted as follows:
If the parameters $\lambda_{\alpha}$, $\lambda_{\sigma}$ are chosen suitably and the noise bound is sufficiently small, then the state $x_t$ converges exponentially to a region around $0$, i.e., the closed loop is practically exponentially stable.
We consider $u^s=0$ (which implies $y^s=0$) for simplicity, but the same result holds qualitatively if $(u^s,y^s)\neq(0,0)$, compare~\cite[Remark 5]{berberich2021guarantees}.
The guaranteed region of attraction $\mathbb{X}_V$ is the sublevel set of a practical Lyapunov function which can be large (i.e., close to the region of attraction of the nominal MPC scheme in Section~\ref{subsec:mpc_nominal}) if $\lambda_{\alpha},\lambda_{\sigma}$ are chosen suitably and $\bar{\varepsilon}$ is sufficiently small.
Similarly, the function $\beta(\bar{\varepsilon})$, i.e., the size of the region to which the closed loop converges, also depends on the parameters $\lambda_{\alpha}$, $\lambda_{\sigma}$, $\bar{\varepsilon}$ and, in particular, it decreases for smaller noise levels $\bar{\varepsilon}$.
As is discussed in more detail in~\cite{berberich2021guarantees}, a larger magnitude of the input $\{u_k^d\}_{k=0}^{N-1}$ generating the data and an increasing length of the data $N$ both improve closed-loop properties under Algorithm~\ref{alg:MPC_rob}, i.e., they increase the region of attraction and decrease the tracking error.
While these findings only reveal qualitative relations between different quantities, it is an important open problem to investigate \emph{quantitative} guidelines for the appropriate selection of parameters in~\eqref{eq:DD_MPC_robust}, which is also analyzed for the example in Section~\ref{sec:simulation}.
To summarize, the MPC scheme based on repeatedly solving Problem~\eqref{eq:DD_MPC_robust} drives the system close to the desired setpoint using a noisy input-output trajectory of finite length.
Sequential system identification and model-based MPC is an obvious alternative to Algorithm~\ref{alg:MPC_rob}.
Advantages of our approach are its simplicity, requiring no prior identification step, while at the same time providing closed-loop guarantees based on noisy data of finite length, which is a challenging problem in identification-based MPC due to the lack of tight estimation error bounds.

\subsection{Data-driven MPC for nonlinear systems}\label{subsec:mpc_nl}
Arguably, one of the biggest challenges in learning-based and data-driven control is the development of methods to control unknown \emph{nonlinear} systems with closed-loop guarantees.
In the following, we address this issue with an MPC scheme based on Theorem~\ref{thm:hankel} which we then apply in the subsequent sections to a practical example.
We do not provide theoretical results for the closed-loop behavior under the presented MPC scheme, which is an issue of our current research.
Let us assume that, instead of~\eqref{eq:sys}, the considered system takes the form
\begin{align}\label{eq:sys_NL}
x_{t+1}&=f(x_t)+g(x_t)u_t,\>\>
y_t=h_0(x_t)+h_1(x_t)u_t
\end{align}
\begin{subequations}\label{eq:DD_MPC_NL}
with unknown vector fields $f$, $g$, $h_0$, $h_1$ of appropriate dimensions.
In the following, our goal is to track a desired output setpoint\footnote{Input setpoints can be included by augmenting the output with a feedthrough term.} $y^T$, i.e., $y_t\to y^T$ for $t\to\infty$, while satisfying input constraints $u_t\in\mathbb{U}$, $t\in\mathbb{I}_{\geq0}$.
To this end, we consider an MPC scheme based on Theorem~\ref{thm:hankel}, similar to the approaches in the previous sections.
In order to account for the nonlinear nature of the dynamics, we update the (noisy) data $\{u_k^d,\tilde{y}_k^d\}_{k=0}^{N-1}$ used for prediction \emph{online} based on current measurements.
In this way, we exploit the fact that the nonlinear system~\eqref{eq:sys_NL} can be locally approximated as a linear system (assuming the vector fields are sufficiently smooth).
Given past $N$ input-output measurements $\{u_k,\tilde{y}_k\}_{k=t-N}^{t-1}$ of~\eqref{eq:sys_NL} at time $t\geq N$, we consider the following open-loop optimal control problem:
\begin{align}
\underset{\substack{\alpha(t),\sigma(t)\\\bar{u}(t),\bar{y}(t)\\u^{s}(t),y^{s}(t)}}{\min}&\sum_{k=0}^{L}
\lVert\bar{u}_k(t)-u^{s}(t)\rVert_R^2+\lVert\bar{y}_k(t)-y^{s}(t)\rVert_Q^2\\\nonumber
&+\lVert y^{s}(t)-y^{\mathrm{T}}\rVert_S^2+\lambda_\alpha\lVert\alpha(t)\rVert_2^2+\lambda_\sigma\lVert\sigma(t)\rVert_2^2\\
\label{eq:DD_MPC_NL_hankel} \text{s.t.}\>\> &\>\begin{bmatrix}
\bar{u}(t)\\\bar{y}(t)+\sigma(t)\end{bmatrix}=\begin{bmatrix}H_{L+n+1}\left(u_{[t-N,t-1]}\right)\\H_{L+n+1}\left(\tilde{y}_{[t-N,t-1]}\right)\end{bmatrix}\alpha(t),\\\label{eq:DD_MPC_NL_init}
&\>\begin{bmatrix}\bar{u}_{[-n,-1]}(t)\\\bar{y}_{[-n,-1]}(t)\end{bmatrix}=\begin{bmatrix}u_{[t-n,t-1]}\\\tilde{y}_{[t-n,t-1]}\end{bmatrix},
\\\label{eq:DD_MPC_NL_TEC}
&\>\begin{bmatrix}\bar{u}_{[L-n,L]}(t)\\\bar{y}_{[L-n,L]}(t)\end{bmatrix}=\begin{bmatrix}u_{n+1}^{s}(t)\\y_{n+1}^{s}(t)\end{bmatrix},\\\label{eq:DD_MPC_NL_sum1}
&\>\sum_{i=0}^{N-L-n-1}\alpha_i(t)=1,\>u^{s}(t)\in\mathbb{U}^{s},\\\label{eq:DD_MPC_NL_constraints}
&\>\bar{u}_k(t)\in\mathbb{U},\>k\in\mathbb{I}_{[0,L]}.
\end{align}
\end{subequations}
The key difference of Problem~\eqref{eq:DD_MPC_NL} to the MPC schemes considered in the previous sections is that the data used for prediction in~\eqref{eq:DD_MPC_NL_hankel} are updated online, thus providing a local linear approximation of the unknown nonlinear system~\eqref{eq:sys_NL}.
Note that~\eqref{eq:DD_MPC_NL} contains a slack variable $\sigma(t)$ as well as regularizations of $\sigma(t)$ and $\alpha(t)$, similar to the robust MPC problem~\eqref{eq:DD_MPC_robust} for LTI systems.
This is due to the fact that the error caused by the local linear approximation of~\eqref{eq:sys_NL} can also be viewed as output measurement noise similar to Section~\ref{subsec:mpc_robust}.

As an additional difference, Problem~\eqref{eq:DD_MPC_NL} includes an artificial setpoint $u^s(t)$, $y^s(t)$ which is optimized online and which enters the terminal equality constraint~\eqref{eq:DD_MPC_NL_TEC}.
The constraint~\eqref{eq:DD_MPC_NL_TEC} is specified over $n+1$ steps such that $(u^s(t),y^s(t))$ is an (approximate) equilibrium of the system and thus, the overall prediction horizon is of length $L+1$.
At the same time, the distance of $y^s(t)$ w.r.t. the actual target setpoint $y^T$ is penalized, where the matrix $S\succ0$ is a design parameter.
The input setpoint $u^s(t)$ lies in some constraint set $\mathbb{U}^s\subseteq\mathrm{int}(\mathbb{U})$.
The idea of optimizing $u^s(t)$, $y^s(t)$ online is inspired by model-based~\cite{limon2008mpc} and data-driven~\cite{berberich2020tracking} tracking MPC, where artificial setpoints can be used to increase the region of attraction or retain closed-loop properties despite online setpoint changes.
In the present problem setting, such an approach has the advantage that, if $S$ is sufficiently small, then the optimal artificial setpoint $(u^{s*}(t),y^{s*}(t))$ appearing in the terminal equality constraint~\eqref{eq:DD_MPC_NL_TEC} remains close to the optimal predicted input-output trajectory $(\bar{u}^*(t),\bar{y}^*(t))$ and hence, close to the initial state $x_t$.
This means that the MPC first drives the system close to the steady-state manifold, where the linearity-based model~\eqref{eq:DD_MPC_NL_hankel} is a good approximation of the nonlinear system dynamics~\eqref{eq:sys_NL} and therefore, the prediction error is small.
Then, the artificial setpoint is slowly shifted towards the target setpoint $y^T$ along the steady-state manifold and hence, the MPC also steers the closed-loop trajectory towards $y^T$.

Finally,~\eqref{eq:DD_MPC_NL_sum1} implies that the weighting vector $\alpha(t)$ sums up to $1$.
The explanation for this modification is that the linearization of~\eqref{eq:sys_NL} at a point which is not a steady-state of~\eqref{eq:sys_NL} generally leads to \emph{affine} (not linear) system dynamics.
Theorem~\ref{thm:hankel} provides a data-driven system parametrization which only applies to linear systems.
In order to parametrize trajectories of an affine system based on measured data, the constraint~\eqref{eq:DD_MPC_NL_sum1} needs to be added since it implies that the constant offset is carried through from the measured data to the predictions.
Problem~\eqref{eq:DD_MPC_NL} can now be applied in a standard receding horizon fashion which is summarized in Algorithm~\ref{alg:MPC_NL}.
%
\begin{algorithm}
\begin{Algorithm}\label{alg:MPC_NL}
\normalfont{\textbf{Nonlinear Data-Driven MPC}}\\
\normalfont{
\textbf{Offline:} Choose upper bound on system order $n$, prediction horizon $L$, cost matrices $Q,R,S\succ0$, regularization parameters $\lambda_{\alpha},\lambda_{\sigma}>0$, constraint sets $\mathbb{U},\mathbb{U}^s$, setpoint $y^T$, and generate data $\{u_k,\tilde{y}_k\}_{k=0}^{N-1}$.}\\
\normalfont{\textbf{Online:}}
\begin{enumerate}
\item At time $t\geq N$, take the past $N$ measurements $\{u_k,\tilde{y}_k\}_{k=t-N}^{t-1}$ and solve~\eqref{eq:DD_MPC_robust}.
\item Apply the input $u_t=\bar{u}_{0}^*(t)$.
\item Set $t=t+1$ and go back to 1).
\end{enumerate}
\end{Algorithm}
\end{algorithm}

It is worth noting that Algorithm~\ref{alg:MPC_NL} only requires solving the strictly convex QP~\eqref{eq:DD_MPC_NL} online, although the underlying control problem involves the nonlinear system~\eqref{eq:sys_NL}.
In this work, we do not address the issue of enforcing that the data $(u,y)$ collected in closed loop and used for prediction in~\eqref{eq:DD_MPC_NL_hankel} are persistently exciting.
It is an obvious practical problem that, upon convergence of the closed loop, the input may eventually be constant and, in particular, not persistently exciting of a sufficient order, which is also an important issue in adaptive MPC~\cite{adetola2011robust}.
For the nonlinear four-tank system investigated in Sections~\ref{sec:simulation} and~\ref{sec:experiment}, we apply the presented MPC without additional modifications enforcing closed-loop persistence of excitation, but we plan to analyze this issue in future research.

\section{Simulation study}\label{sec:simulation}

In this section, we apply the MPC scheme for nonlinear systems discussed in Section~\ref{subsec:mpc_nl} to a simulation model of the four-tank system originally considered in~\cite{raff2006nonlinear}.
The continuous-time system dynamics can be described as
\begin{align}\label{eq:four_tank_sim}
\dot{x}_1&=-\frac{a_1}{A_1}\sqrt{2gx_1}+\frac{a_3}{A_1}\sqrt{2gx_3}+\frac{\gamma_1}{A_1}u_1,\\\nonumber
\dot{x}_2&=-\frac{a_2}{A_2}\sqrt{2gx_2}+\frac{a_4}{A_2}\sqrt{2gx_4}+\frac{\gamma_2}{A_2}u_2,\\\nonumber
\dot{x}_3&=-\frac{a_3}{A_3}\sqrt{2gx_3}+\frac{1-\gamma_2}{A_3}u_2,\\\nonumber
\dot{x}_4&=-\frac{a_4}{A_4}\sqrt{2gx_4}+\frac{1-\gamma_1}{A_4}u_1,
\end{align}
where $x_i$ is the water level of tank $i$ in $\text{cm}$, $u_i$ the flow rate of pump $i$ in $\text{cm}^3/s$, and the other terms are system parameters, whose values are taken from~\cite{raff2006nonlinear} and summarized in Table~\ref{tab:sim_model_param}.
\begin{table}
\begin{center}
\begin{tabular}{c|c|c|c}
$A_1=A_2$\normalfont{:} & $A_3=A_4$\normalfont{:} & $a_1$\normalfont{:} & $a_2$\normalfont{:}\\
$50.27\text{\normalfont{cm}}^2$ & $28.27\text{\normalfont{cm}}^2$ & $0.233\text{\normalfont{cm}}^2$ & $0.242\text{\normalfont{cm}}^2$\\\hline
$a_3=a_4$\normalfont{:} & $\gamma_1=\gamma_2$\normalfont{:} & $g$\normalfont{:}\\
$0.127\text{\normalfont{cm}}^2$ & $0.4$ & $981\text{\normalfont{cm}}^2/\text{\normalfont{s}}$\\
\end{tabular}
\vskip6.5pt
\caption{Parameter values of the simulation model~\eqref{eq:four_tank_sim}.}\label{tab:sim_model_param}
\end{center}
\end{table}
The output of the system is given by $y=\begin{bmatrix}x_1&x_2\end{bmatrix}^\top$.
For the following simulation study, we assume that this output can be measured exactly without noise since this allows us to better investigate and illustrate the interplay between the nonlinear system dynamics and suitable design parameters of Problem~\eqref{eq:DD_MPC_NL} leading to a good closed-loop operation.
In Section~\ref{sec:experiment}, we show that the proposed MPC scheme is also applicable in a real-world experiment with noisy measurements.

\begin{figure}
		\begin{center}
		\subfigure
		{\includegraphics[width=0.49\textwidth]{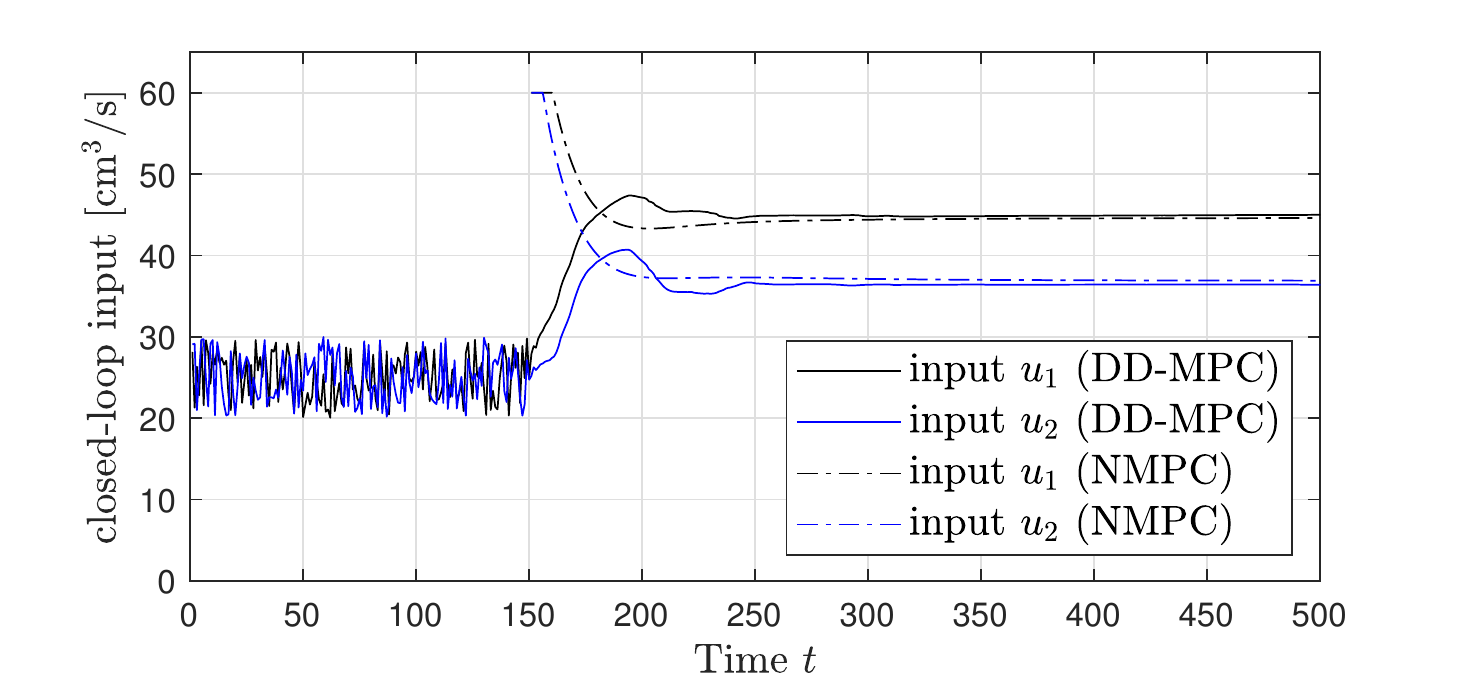}}
		\subfigure
		{\includegraphics[width=0.49\textwidth]{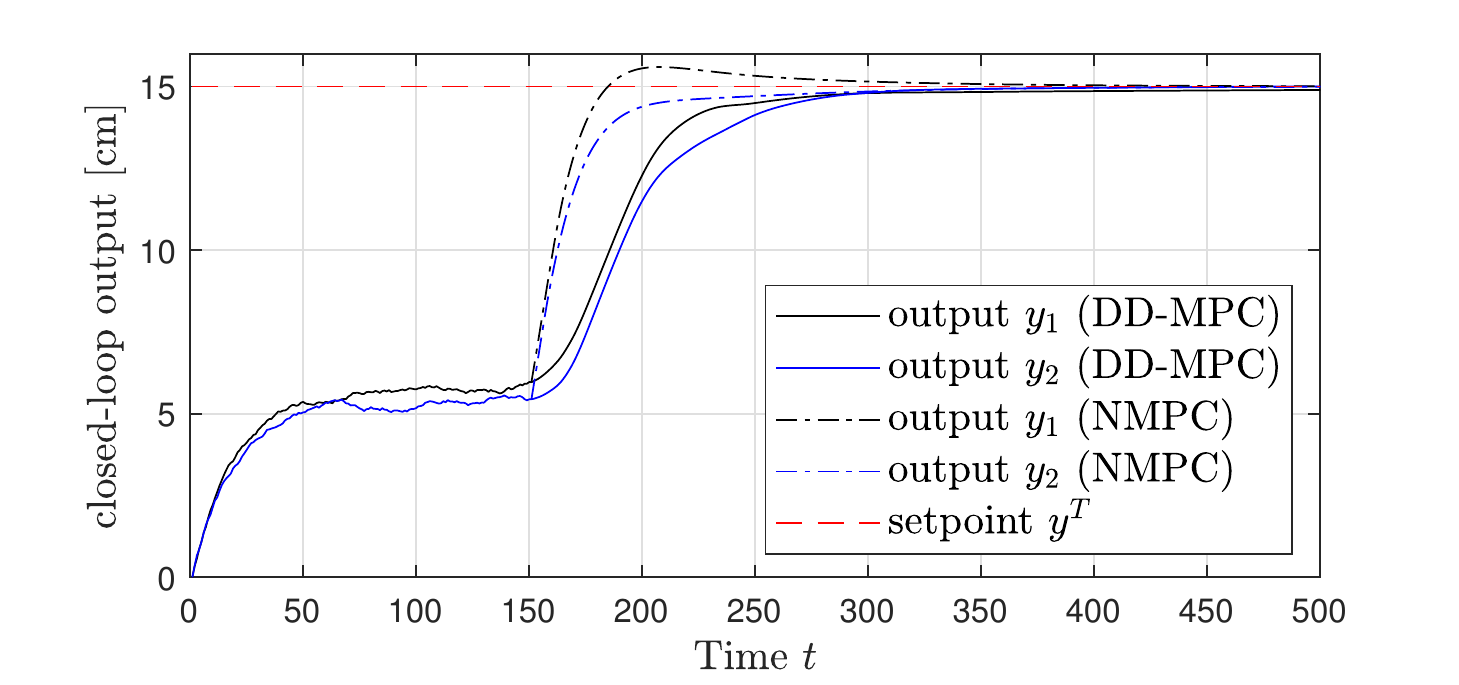}}
		\end{center}
		\caption{Closed-loop input-output trajectory, resulting from data-driven MPC (DD-MPC, Algorithm~\ref{alg:MPC_NL}) and model-based nonlinear MPC (NMPC,~\cite{koehler2020nonlinear}) to the four-tank system in simulation.}	\label{fig:simulation_io}
\end{figure}

\begin{figure}
\begin{center}
\includegraphics[width=0.5\textwidth]{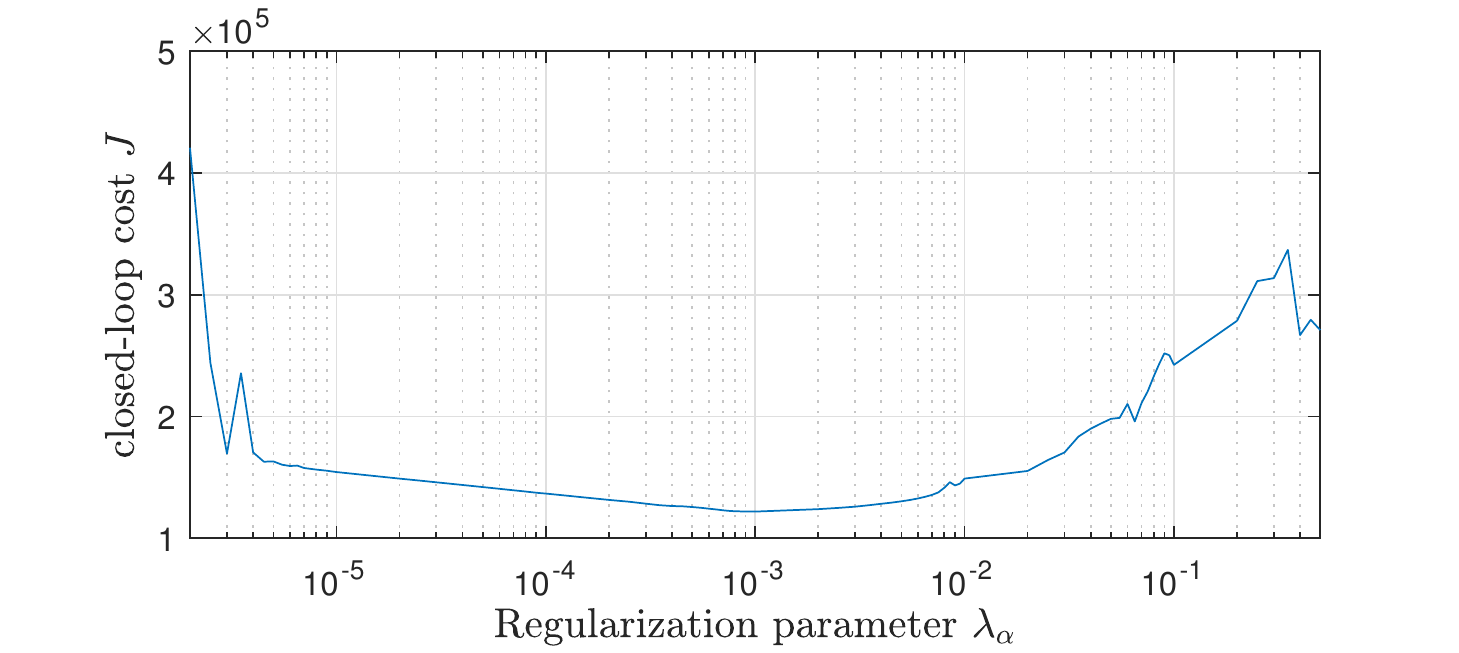}
\end{center}
\caption{Closed-loop cost $J$ depending on the parameter $\lambda_{\alpha}$.}
\label{fig:cost_lambda_alpha}
\end{figure}

We now apply the nonlinear MPC scheme from Section~\ref{subsec:mpc_nl} (compare Algorithm~\ref{alg:MPC_NL}) to the discrete-time nonlinear system obtained via Euler discretization with sampling time $T_s=1.5$ seconds of~\eqref{eq:four_tank_sim}.
Our goal is to track the setpoint $y^T=\begin{bmatrix}15&15\end{bmatrix}^\top$ while satisfying the input constraints $u_t\in\mathbb{U}=[0,60]^2$.
To this end, we apply an input sequence sampled uniformly from\footnote{This interval is chosen sufficiently large and does not contain zero due to the fact that too small inputs imply that the outputs are also small and thus lie in a region where the sensors of the experimental setup in Section~\ref{sec:experiment} are less accurate.} $u_k\in[20,30]^2$ over the first $N$ time steps to collect initial data, where the system is initialized at $x_0=0$.
Thereafter, for each $t\geq N$, we solve Problem~\eqref{eq:DD_MPC_NL}, apply the first component of the optimal predicted input, and update the data $\{u_k,y_k\}_{k=t-N}^{t-1}$ used for prediction in~\eqref{eq:DD_MPC_NL_hankel} in the next time step based on the current measurements.
We use the parameters
\begin{align}\label{eq:four_tank_sim_param}
&N=150,\>\>L=35,\>\>Q=I,\>\>R=2I,\\\nonumber
&S=20 I,\>\>\lambda_{\alpha}=5\cdot 10^{-5},\>\>\lambda_{\sigma}=2\cdot10^5,
\end{align}
and we choose the equilibrium input constraints as $\mathbb{U}^s=[0.6,59.4]^2$.
Further, the value of $n$ used in~\eqref{eq:DD_MPC_NL} (i.e., our estimate of the system order) is chosen as $3$.
This suffices for the application of data-driven MPC since the lag of (the linearization of) the above system is $2$ and the implicit prediction model remains valid as long as $n$ is an upper bound on the lag (compare~\cite{markovsky2008data} for details).
The closed-loop input and output trajectories under the MPC scheme with these parameters can be seen in Fig.~\ref{fig:simulation_io}.
After the initial excitation phase $t\in\mathbb{I}_{[0,N-1]}$, the MPC successfully steers the output to the desired target setpoint.
First, we note that updating the data used for prediction in~\eqref{eq:DD_MPC_NL_hankel} is a crucial ingredient of our MPC approach for nonlinear systems.
In particular, if we do not update the data online but only use the first $N$ input-output measurements $\{u_k,y_k\}_{k=0}^{N-1}$ for prediction, then the closed loop does not converge to the desired output $y^T$ and instead yields a significant permanent offset due to the model mismatch.
For comparison, Fig.~\ref{fig:simulation_io} also shows the closed-loop trajectory starting at time $t=N$ resulting from a nonlinear tracking MPC scheme with full model knowledge and state measurements from~\cite{koehler2020nonlinear}, where the parameters are as above except for $S=200I$ and $R=0.1I$.
The two MPC schemes exhibit similar convergence speed although the data-driven MPC uses ``less aggressive'' parameters due to the slack variable $\sigma(t)$ which implicitly relaxes the terminal equality constraint~\eqref{eq:DD_MPC_NL_TEC}.
It has been observed in the literature, e.g.,~\cite{elokda2019quadcopters}, that the choice of the regularization parameter $\lambda_{\alpha}$ has an essential impact on the closed-loop performance of data-driven MPC.
In the following, we investigate in more detail how the specific choice of $\lambda_{\alpha}$ influences the closed-loop performance.
To this end, we perform closed-loop simulations for a range of values $\lambda_{\alpha}$ and, for each of these simulations, we compute the corresponding cost as the deviation of the closed-loop output from the target setpoint $y^T$, i.e., $J=\sum_{t=N}^{500}\lVert y_t-y^T\rVert_S^2$.
For comparison, we note that the parameters in~\eqref{eq:four_tank_sim_param} lead to a closed-loop cost of $J=1.42\cdot10^5$, whereas the model-based nonlinear MPC shown in Fig.~\ref{fig:simulation_io} leads to $J=3.1\cdot10^4$.
Fig.~\ref{fig:cost_lambda_alpha} shows the closed-loop cost depending on the parameter $\lambda_{\alpha}$ with all other parameters as in~\eqref{eq:four_tank_sim_param}.
Although the cost strongly depends on $\lambda_{\alpha}$, it can be seen that a wide range of values $\lambda_{\alpha}\in[2\cdot10^{-5},0.01]$ leads to a good performance, i.e., $J\leq 1.5\cdot10^5$.
If $\lambda_{\alpha}$ is chosen too small, then the robustness w.r.t. the nonlinearity deteriorates and the influence of numerical inaccuracies increases, which leads to a cost increase.
This is in accordance with Theorem~\ref{thm:DD_MPC_rob} which requires that $\lambda_{\alpha}$ is suitably chosen (in particular, it cannot be arbitrarily small).
On the other hand, if $\lambda_{\alpha}$ is chosen too large then the closed-loop cost increases significantly since too small choices of the vector $\alpha(t)$ shift the input and output to which the closed loop converges towards zero, i.e., large values of $\lambda_{\alpha}$ increase the asymptotic tracking error.
To summarize, since a wide range of values $\lambda_{\alpha}$ leads (approximately) to the minimum achievable cost, tuning the parameter $\lambda_{\alpha}$ is easy for the present example.

Next, we analyze how different choices of other design parameters influence the closed-loop cost.
Table~\ref{tab:simulation_param} displays ranges for various parameters for which the cost $J$ is less than $1.5\cdot10^5$, when keeping all other parameters as in~\eqref{eq:four_tank_sim_param}.
The data length $N$ needs to be sufficiently large such that the input is persistently exciting, but choosing it too large deteriorates the performance since then the data used for prediction in~\eqref{eq:DD_MPC_NL_hankel} cover a larger region of the state-space and the implicit linearity-based ``model'' is a less accurate approximation of the nonlinear dynamics~\eqref{eq:four_tank_sim}.
This is in contrast to the results on robust data-driven MPC for linear systems in Section~\ref{subsec:mpc_robust}, where larger data lengths always improve the closed-loop performance (cf.~\cite{berberich2021guarantees}).
Similarly, too large values for the prediction horizon $L$ are detrimental since they imply that the predicted trajectories are further away from the initial state, where the prediction accuracy deteriorates.
On the other hand, too short horizons $L$ lead to worse robustness due to the terminal equality constraints~\eqref{eq:DD_MPC_NL_TEC}.
The assumed system order $n$ cannot be larger than $4$ due to the dependence of the required persistence of excitation on $n$ and since larger values of $n$ effectively shorten the prediction horizon due to the terminal equality constraints~\eqref{eq:DD_MPC_NL_TEC}, which are specified over $n+1$ time steps.
If $N$ and $L$ are increased to $N=190$ and $L=40$, then the closed-loop output still converges to $y^T$, e.g., for the upper bound $10$ on the system order.
\begin{table}
\begin{center}
\begin{tabular}{c|c|c}
$N$\normalfont{:} & $L$\normalfont{:} & \normalfont{assumed system order:}\\
$\mathbb{I}_{[130,159]}$&
$\mathbb{I}_{[32,41]}$&
$\mathbb{I}_{[2,4]}$\\\hline
$\bar{s}$\normalfont{:} & $\lambda_{\alpha}$\normalfont{:} & $\lambda_{\sigma}$\normalfont{:}\\
$[16,3\cdot 10^2]$&
$[2\cdot10^{-5},0.01]$&
$[4\cdot10^2,10^6]$\\
\end{tabular}
\vskip6.5pt
\caption{Parameters leading to a closed-loop cost $J\leq1.5\cdot10^{5}$.}\label{tab:simulation_param}
\end{center}
\end{table}

Further, Table~\ref{tab:simulation_param} displays values of $\bar{s}$ leading to a good closed-loop performance if the matrix $S$ is chosen as $S=\bar{s} I$.
The value $\bar{s}$ cannot be arbitrarily large since it needs to be small enough such that the artificial setpoint $(u^s(t),y^s(t))$ and therefore the predicted trajectories remain close to the initial state, where the prediction accuracy of the data-dependent model~\eqref{eq:DD_MPC_NL_hankel} is acceptable (compare the discussion in Section~\ref{subsec:mpc_nl}).
On the other hand, for too small values of $\bar{s}$, the asymptotic tracking error increases since the artificial steady-state is close to the initial condition and thus, the regularization of $\alpha$ w.r.t. zero dominates the cost of~\eqref{eq:DD_MPC_NL}.
Moreover, the parameter $\lambda_{\sigma}$ can be chosen in a relatively large range.
To summarize, the MPC scheme shown in Section~\ref{subsec:mpc_nl} can successfully control the nonlinear four-tank system from~\cite{raff2006nonlinear} in simulation, and the influence of system and design parameters on the closed-loop performance confirms our theoretical findings.

\section{Experimental application}\label{sec:experiment}
\begin{figure}
		\begin{center}
		\subfigure
		{\includegraphics[width=0.49\textwidth]{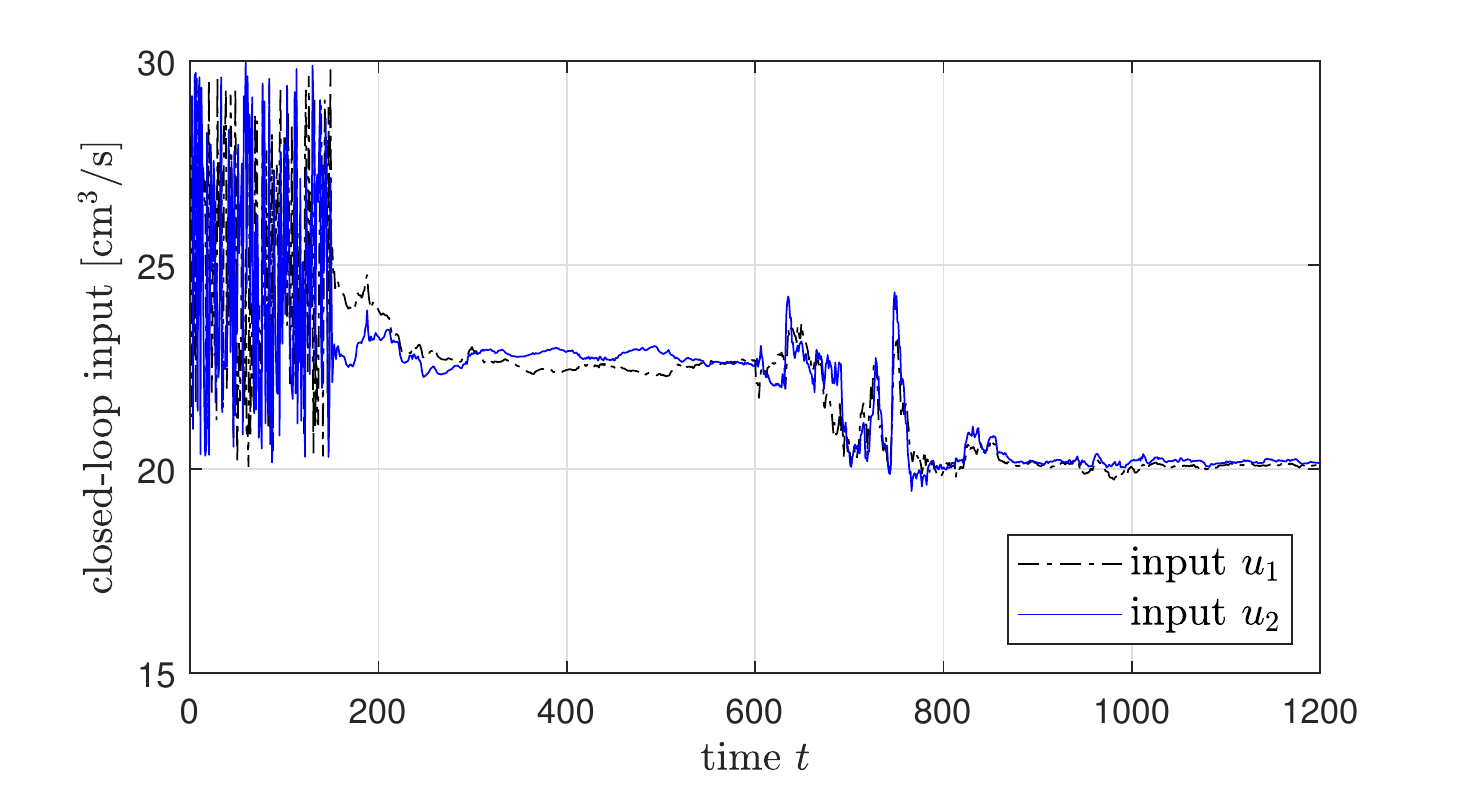}}
		\subfigure
		{\includegraphics[width=0.49\textwidth]{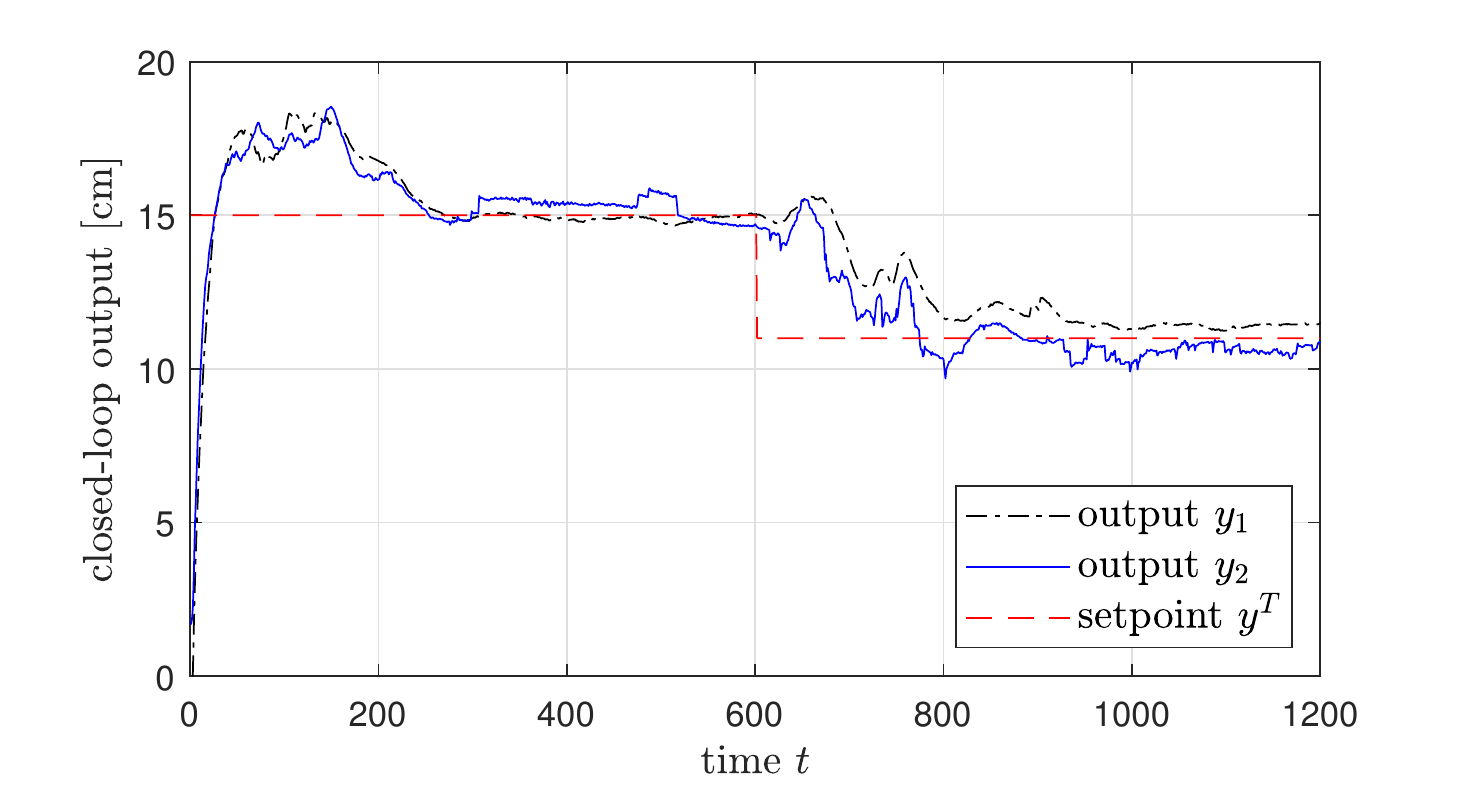}}
		\end{center}
		\caption{Closed-loop input-output trajectory, resulting from the application of the data-driven MPC scheme presented in Section~\ref{subsec:mpc_nl} to the four-tank system in an experiment.}	\label{fig:experiment_io}
\end{figure}

In the following, we apply the MPC scheme presented in Section~\ref{subsec:mpc_nl} in an experimental setup to the four-tank system by Quanser.
This system possesses qualitatively the same dynamics as~\eqref{eq:four_tank_sim}, but the parameter values differ (compare~\cite{quanser2021coupled} for details).
Nevertheless, as we show in the following, the presented nonlinear data-driven MPC scheme can successfully control the system using the same design parameters as in Section~\ref{sec:simulation} due to its ability to adapt to changing operating conditions, in particular by updating the data used for prediction online.
We use the same sampling time $T_s=1.5$ seconds as in Section~\ref{sec:simulation}.
Similar to Section~\ref{sec:simulation}, we first apply an open-loop input sampled uniformly from $u_k\in[20,30]^2$ in order to generate data of length $N=150$.
Thereafter, we compute the input applied to the plant via an MPC scheme based on Problem~\eqref{eq:DD_MPC_NL}, where the design parameters are chosen exactly as in Section~\ref{sec:simulation}, i.e., as in~\eqref{eq:four_tank_sim_param}.
In addition to only tracking the setpoint $y^T=\begin{bmatrix}15&15\end{bmatrix}^\top$ in the time interval $t\in\mathbb{I}_{[0,600]}$, we include an online setpoint change for the time interval $t\in\mathbb{I}_{[601,1200]}$ to $y^T=\begin{bmatrix}11&11\end{bmatrix}^\top$.
We note that the computation time for solving the strictly convex QP~\eqref{eq:DD_MPC_NL} is negligible compared to the sampling time of $1.5$ seconds.
The resulting closed-loop input-output trajectory is displayed in Fig.~\ref{fig:experiment_io}.
After the initial exploration phase of length $N$, the closed-loop output first converges towards the setpoint $\begin{bmatrix}15&15\end{bmatrix}^\top$ and after time $t=600$, the output converges towards the second setpoint $\begin{bmatrix}11&11\end{bmatrix}^\top$, i.e., the MPC approximately solves our control problem.
Similar to the simulation results in Section~\ref{sec:simulation}, the closed loop has a large steady-state tracking error if at all times only the first $N=150$ data points are used for prediction, underpinning the importance of updating the measured data in~\eqref{eq:DD_MPC_NL_hankel} online when controlling nonlinear systems.
However, Fig.~\ref{fig:experiment_io} also illustrates a drawback of the presented approach which always relies on the last $N$ input-output measurements.
Upon convergence, the closed-loop input is approximately constant and, although the \emph{qualitative} persistence of excitation condition in Definition~\ref{def:pe} is still fulfilled, some of the singular values of the input Hankel matrix are very small, which deteriorates the prediction accuracy and hence the closed-loop performance (compare also the discussion at the end of Section~\ref{subsec:mpc_nl}).
Therefore, the closed-loop output does not exactly converge to the setpoint but oscillates within a small region around $y^T$.
Moreover, when the setpoint change is initiated at time $t=600$, the past $N=150$ input-output data points contain only little information about the system behavior, which deteriorates the transient closed-loop behavior.
It is possible to overcome these issues, e.g., by stopping the data updates after the setpoint is reached or by explicitly enforcing closed-loop persistence of excitation.
We plan to investigate the benefit of such measures in future research.

Comparing Figures~\ref{fig:simulation_io} and~\ref{fig:experiment_io}, we observe an important advantage of the presented MPC framework.
Clearly, the two four-tank systems~\cite{raff2006nonlinear} and~\cite{quanser2021coupled} have different parameters, e.g., the steady-state inputs leading to the output $y^T$ differ significantly.
In particular, the model~\eqref{eq:four_tank_sim} does not accurately describe the four-tank system~\cite{quanser2021coupled}, e.g., due to differing pump flow rates, differing tube diameters, manufacturing inaccuracies, aging, and since the model~\eqref{eq:four_tank_sim} is not even an \emph{exact} representation of the physical reality for the four-tank system considered in~\cite{raff2006nonlinear}.
In order to implement a (nonlinear) model-based MPC as in~\cite{raff2006nonlinear}, all of the mentioned quantities need to be carefully modeled which can be a challenging and time-consuming task.
On the other hand, estimating an accurate model based on an open-loop experiment is also difficult due to the nonlinear nature of~\eqref{eq:four_tank_sim} and since only input-output measurements are available, see, e.g.,~\cite{calliess2014conservative}.
In contrast, the proposed MPC leads to an acceptable closed-loop performance without any modifications compared to the simulation in Section~\ref{sec:simulation} due to the fact that it naturally adapts to the operating conditions.
This makes our MPC framework both very simple to apply, since no modeling or nonlinear identification tasks need to be carried out, and reliable, since the framework allows for rigorous theoretical guarantees (although so far only for linear systems).

\section{Conclusion}\label{sec:conclusion}

We presented an MPC framework to control unknown systems using only measured data.
We discussed simple MPC schemes for LTI systems which admit strong theoretical guarantees in closed loop both with and without measurement noise.
Further, we proposed a modification which can be used to control unknown nonlinear systems by repeatedly updating the data used for prediction and exploiting local linear approximations.
Finally, we applied this approach in simulation and in an experiment to a nonlinear four-tank system.
Important advantages of the presented framework are its simplicity, the fact that no explicit model knowledge is required, the low computational complexity (solving a QP), the possibility to adapt to online changes in the system dynamics, and the applicability to (unknown) nonlinear systems.
In particular, obtaining accurate models of nonlinear systems using noisy input-output data is a very challenging and largely open research problem.
On the other hand, the presented framework admits desirable theoretical guarantees for LTI systems, and analogous results for nonlinear systems are the subject of our current research.
Another interesting direction for future research is the practical and theoretical comparison to MPC based on (online) system identification, e.g.,~\cite{adetola2011robust,nguyen2020output}.

\begin{funding}
This work was funded by Deutsche Forschungsgemeinschaft (DFG, German Research Foundation) under Germany’s Excellence Strategy - EXC 2075 - 390740016 and the International Research Training Group Soft Tissue Robotics (GRK 2198/1 - 277536708).
This project has received funding from the European Research Council (ERC) under the European Union’s Horizon 2020 research and innovation programme (grant agreement No 948679).
The authors thank the International Max Planck Research School for Intelligent Systems (IMPRS-IS)
for supporting Julian Berberich.
\end{funding}

\bibliographystyle{plain}
\bibliography{Literature}  

\begin{thebibliography}{10}

\bibitem{quanser2021coupled}
Quanser coupled tanks system data sheet.
\newblock [Online] {\url{https://www.quanser.com/products/coupled-tanks/}}.
\newblock Accessed: 2021-01-21.

\bibitem{adetola2011robust}
V.~Adetola and M.~Guay.
\newblock Robust adaptive {MPC} for constrained uncertain nonlinear systems.
\newblock {\em Int. J. Adaptive Control and Signal Processing}, 25(2):155--167,
  2011.

\bibitem{aswani2013provably}
A.~Aswani, H.~Gonzalez, S.~S. Sastry, and C.~Tomlin.
\newblock Provably safe and robust learning-based model predictive control.
\newblock {\em Automatica}, 49(5):1216--1226, 2013.

\bibitem{berberich2020trajectory}
J.~Berberich and F.~Allg\"ower.
\newblock A trajectory-based framework for data-driven system analysis and
  control.
\newblock In {\em Proc. European Control Conf.}, pages 1365--1370, 2020.

\bibitem{berberich2020tracking}
J.~Berberich, J.~K{\"o}hler, M.~A. M{\"u}ller, and F.~Allg{\"o}wer.
\newblock Data-driven tracking {MPC} for changing setpoints.
\newblock In {\em Proc. IFAC World Congress}, 2020.
\newblock to appear.

\bibitem{berberich2020constraints}
J.~Berberich, J.~K{\"o}hler, M.~A. M{\"u}ller, and F.~Allg{\"o}wer.
\newblock Robust constraint satisfaction in data-driven {MPC}.
\newblock In {\em Proc. Conf. Decision and Control}, pages 1260--1267, 2020.

\bibitem{berberich2021guarantees}
J.~Berberich, J.~K{\"o}hler, M.~A. M{\"u}ller, and F.~Allg{\"o}wer.
\newblock Data-driven model predictive control with stability and robustness
  guarantees.
\newblock {\em IEEE Transactions on Automatic Control}, 66(4):1702--1717, 2021.

\bibitem{berberich2021on}
J.~Berberich, J.~K{\"o}hler, M.~A. M{\"u}ller, and F.~Allg{\"o}wer.
\newblock On the design of terminal ingredients for data-driven {MPC}.
\newblock {\em arXiv:2101.05573}, 2021.

\bibitem{bongard2021robust}
J.~Bongard, J.~Berberich, J.~K{\"o}hler, and F.~Allg{\"o}wer.
\newblock Robust stability analysis of a simple data-driven model predictive
  control approach.
\newblock {\em arXiv:2103.00851}, 2021.

\bibitem{calliess2014conservative}
J.-P. Calliess.
\newblock {\em Conservative decision-making and inference in uncertain
  dynamical systems}.
\newblock PhD thesis, University of Oxford, 2014.

\bibitem{coulson2019deepc}
J.~Coulson, J.~Lygeros, and F.~D{\"o}rfler.
\newblock Data-enabled predictive control: in the shallows of the {DeePC}.
\newblock In {\em Proc. European Control Conf.}, pages 307--312, 2019.

\bibitem{coulson2020distributionally}
J.~Coulson, J.~Lygeros, and F.~D{\"o}rfler.
\newblock Distributionally robust chance constrained data-enabled predictive
  control.
\newblock {\em arXiv:2006.01702}, 2020.

\bibitem{elokda2019quadcopters}
E.~Elokda, J.~Coulson, J.~Lygeros, and F.~D{\"o}rfler.
\newblock Data-enabled predictive control for quadcopters.
\newblock {\em ETH Zurich, Research Collection:10.3929/ethz-b-000415427}, 2019.

\bibitem{hewing2020learning}
L.~Hewing, K.~P. Wabersich, M.~Menner, and M.~N. Zeilinger.
\newblock Learning-based model predictive control: Toward safe learning in
  control.
\newblock {\em Ann. Rev. Control, Robotics, and Autonomous Systems},
  3:269--296, 2020.

\bibitem{hou2013model}
Z.-S. Hou and Z.~Wang.
\newblock From model-based control to data-driven control: Survey,
  classification and perspective.
\newblock {\em Information Sciences}, 235:3--35, 2013.
\newblock 

\bibitem{koehler2020nonlinear}
J.~K\"ohler, M.~A. M\"uller, and F.~Allg\"ower.
\newblock A nonlinear tracking model predictive control scheme for dynamic
  target signals.
\newblock {\em Automatica}, 118:109030, 2020.

\bibitem{limon2008mpc}
D.~Lim{\'o}n, I.~Alvarado, T.~Alamo, and E.~F. Camacho.
\newblock {MPC} for tracking piecewise constant references for constrained
  linear systems.
\newblock {\em Automatica}, 44(9):2382--2387, 2008.

\bibitem{markovsky2008data}
I.~Markovsky and P.~Rapisarda.
\newblock Data-driven simulation and control.
\newblock {\em Int. J. Control}, 81(12):1946--1959, 2008.

\bibitem{nguyen2020output}
T.~W. Nguyen, S.~A.~U. Islam, A.~L. Bruce, A.~Goel, D.~S. Bernstein, and I.~V.
  Kolmanovsky.
\newblock Output-feedback {RLS}-based model predictive control.
\newblock In {\em Proc. American Control Conf.}, pages 2395--2400, 2020.

\bibitem{raff2006nonlinear}
T.~Raff, S.~Huber, Z.~K. Nagy, and F.~Allg{\"o}wer.
\newblock Nonlinear model predictive control of a four tank system: An
  experimental stability study.
\newblock In {\em Proc. Int. Conf. Control Applications}, pages 237--242, 2006.

\bibitem{rawlings2017model}
J.~B. Rawlings, D.~Q. Mayne, and M.~M. Diehl.
\newblock {\em Model Predictive Control: Theory, Computation, and Design}.
\newblock Nob Hill Pub, 2nd edition, 2017.

\bibitem{willems2005note}
J.~C. Willems, P.~Rapisarda, I.~Markovsky, and B.~{De Moor}.
\newblock A note on persistency of excitation.
\newblock {\em Systems \& Control Letters}, 54:325--329, 2005.

\bibitem{yang2015data}
H.~Yang and S.~Li.
\newblock A data-driven predictive controller design based on reduced hankel
  matrix.
\newblock In {\em Proc. Asian Control Conf.}, pages 1--7, 2015.

\end{thebibliography}

\begin{wrapfigure}{l}{25mm} 
    \includegraphics[width=1in,height=1.25in,clip,keepaspectratio]{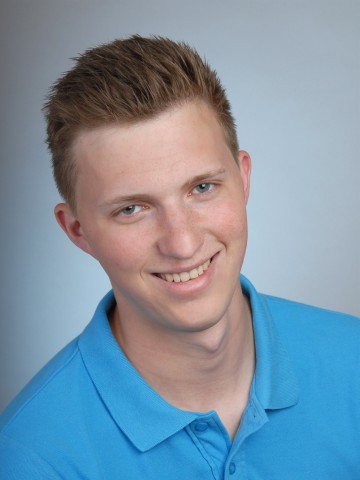}
  \end{wrapfigure}\par
  \textbf{Julian Berberich} 
received the Master’s degree in Engineering Cybernetics from the University of Stuttgart, Germany, in 2018. Since 2018, he has been a Ph.D. student at the Institute for Systems Theory and Automatic Control under supervision of Prof. Frank Allg\"ower and a member of the International Max-Planck Research School (IMPRS). He has received the Outstanding Student Paper Award at the 59th Conference on Decision and Control in 2020. His research interests are in the area of data-driven system analysis and control.\par
\vskip10pt

\begin{wrapfigure}{l}{25mm} 
    \includegraphics[width=1in,height=1.25in,clip,keepaspectratio]{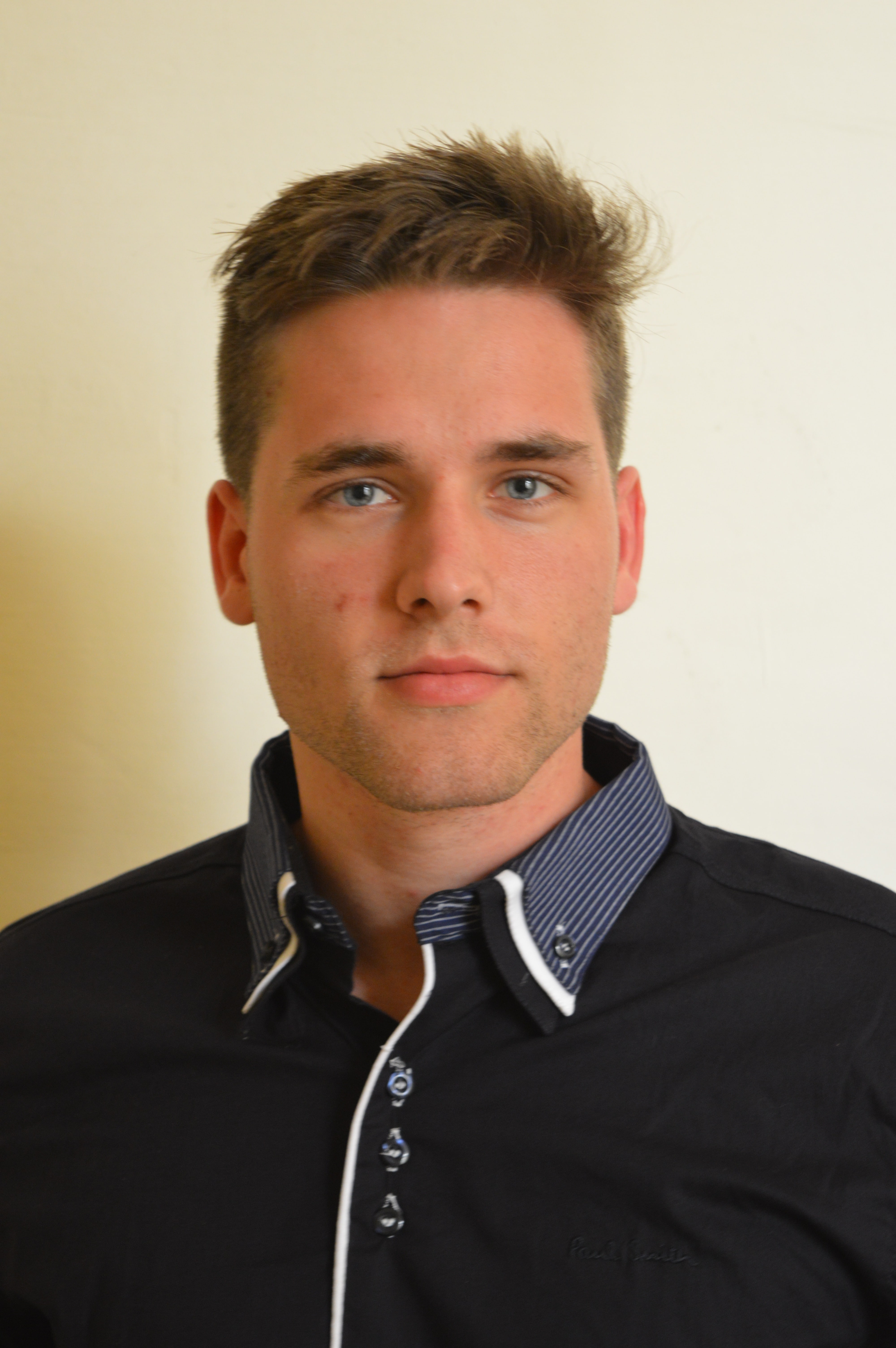}
  \end{wrapfigure}\par
  \textbf{Johannes K\"ohler}
 received his Master degree in Engineering Cybernetics from the University of Stuttgart, Germany, in 2017.
He has since been a doctoral student at the \emph{Institute for Systems Theory and Automatic Control} under the supervision of Prof. Frank Allg\"ower and a member of the Graduate School Soft Tissue Robotics at the University of Stuttgart. 
His research interests are in the area of model predictive control.\par
\vskip10pt

\begin{wrapfigure}{l}{25mm} 
    \includegraphics[width=1in,height=1.25in,clip,keepaspectratio]{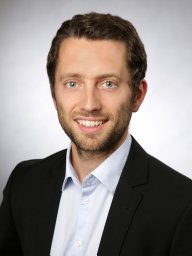}
  \end{wrapfigure}\par
  \textbf{Matthias A. M\"uller}
  received a Diploma degree in Engineering Cybernetics from the University of Stuttgart, Germany, and an M.S. in Electrical and Computer Engineering from the University of Illinois at Urbana-Champaign, US, both in 2009.
In 2014, he obtained a Ph.D. in Mechanical Engineering, also from the University of Stuttgart, Germany, for which he received the 2015 European Ph.D. award on control for complex and heterogeneous systems. Since 2019, he is director of the Institute of Automatic Control and full professor at the Leibniz University Hannover, Germany. 
He obtained an ERC Starting Grant in 2020 and is recipient of the inaugural Brockett-Willems Outstanding Paper Award for the best paper published in Systems \& Control Letters in the period 2014-2018. His research interests include nonlinear control and estimation, model predictive control, and data-/learning-based control, with application in different fields including biomedical engineering.\par
\vskip10pt

\begin{wrapfigure}{l}{25mm} 
    \includegraphics[width=1in,height=1.25in,clip,keepaspectratio]{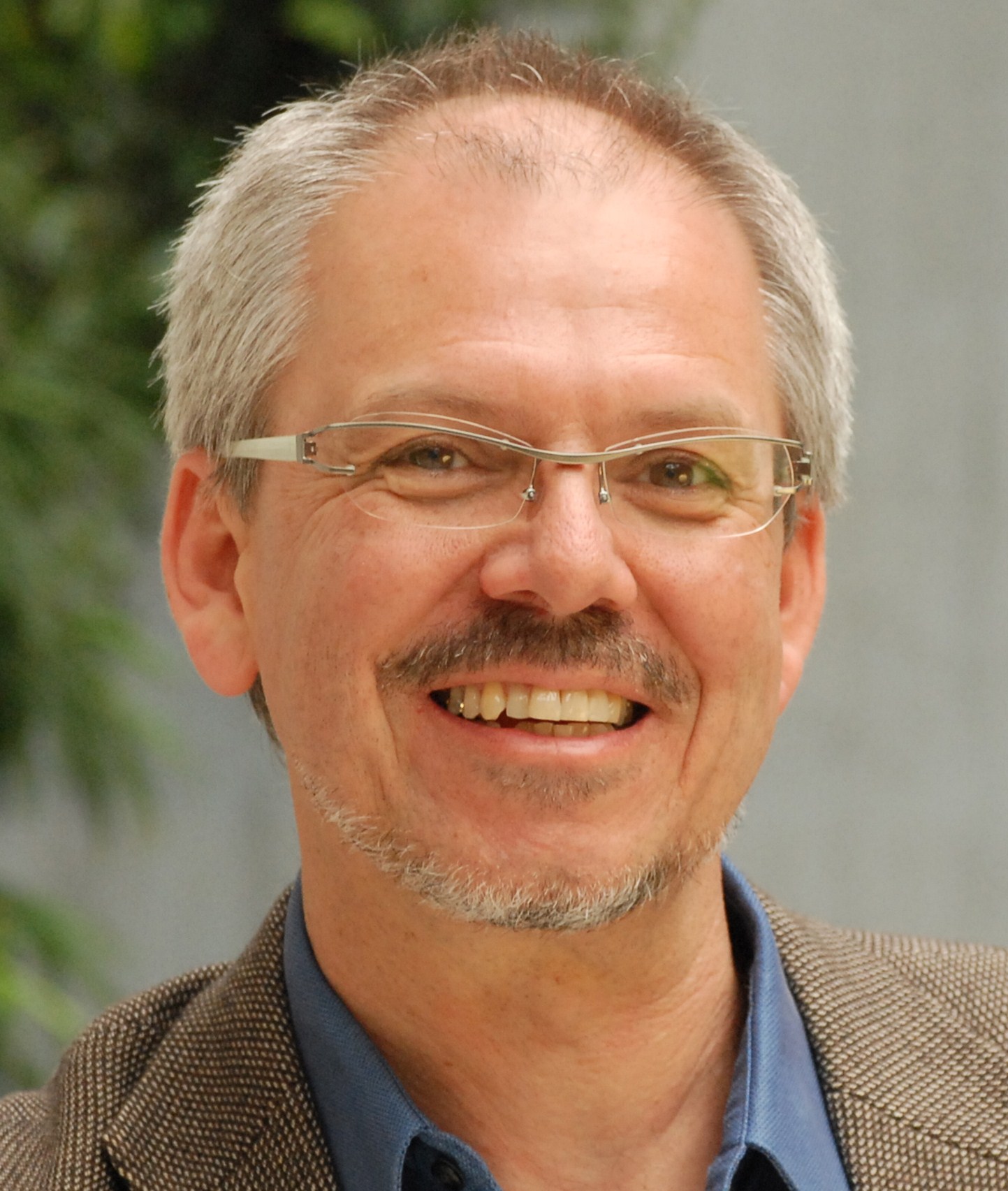}
  \end{wrapfigure}\par
  \textbf{Frank Allg\"ower}
is professor of mechanical engineering at the University of Stuttgart, Germany, and Director of the Institute for Systems Theory and Automatic Control (IST) there.\\ 
Frank is active in serving the community in several roles: Among others he has been President of the International Federation of Automatic Control (IFAC) for the years 2017-2020, Vice-president for Technical Activities of the IEEE Control Systems Society for 2013/14, and Editor of the journal Automatica from 2001 until 2015. From 2012 until 2020 Frank served in addition as Vice-president for the German Research Foundation (DFG), which is Germany’s most important research funding organization. \\
His research interests include predictive control, data-based control, networked control, cooperative control, and nonlinear control with application to a wide range of fields including systems biology.\par

\end{document}